\documentclass[notitlepage,12pt]{article}

\usepackage{a4wide}
\usepackage{graphicx}

  \renewenvironment{thebibliography}[1]{%
    \begin{oldthebibliography}{#1}
      \setlength{\itemsep}{0.9ex}%
  }%
  {%
    \end{oldthebibliography}%
  }

\begin{document}
\renewcommand{\refname}{Publications}

\begin{titlepage}

\newlength{\Size}
\setlength{\Size}{0.2\textwidth}
\newlength{\Shift}
\settoheight{\Shift}{L}
\addtolength{\Shift}{-\Size}


\vspace{3cm}
\begin{center}
\bf \Huge John Clifford Barton\\
\Large 1923 -- 2002

\begin{picture}(430,400)
\put(0,-20)
{\includegraphics[width=15cm]{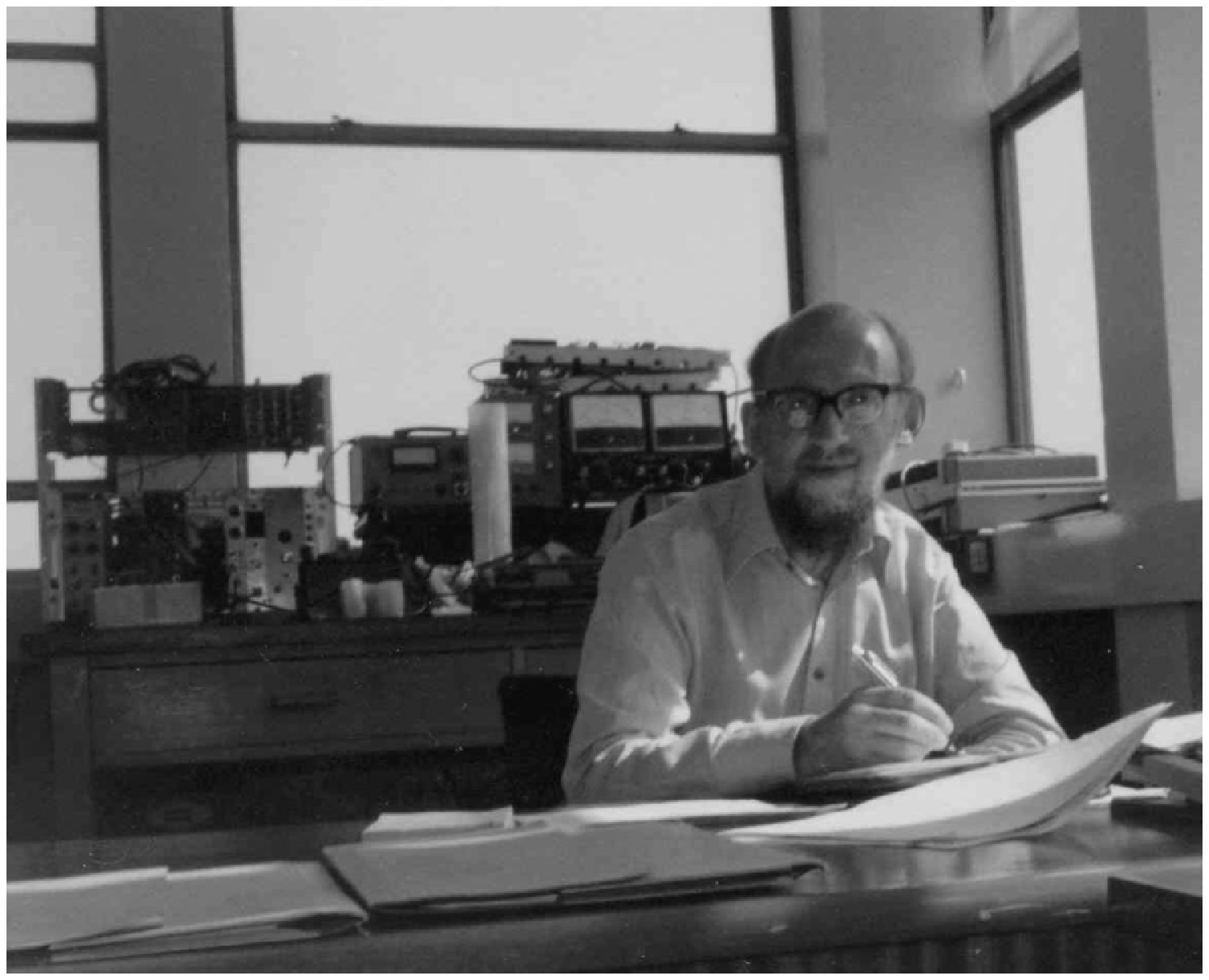}}
\end{picture}
\vspace{2cm}

\Large{Obituary and Bibliography}
\end{center}

\normalsize\rm 
\begin{center}

\end{center} 

\end{titlepage}

\suppressfloats[t]   

\section*{John Clifford Barton}

John Barton was one of the founders of particle
astrophysics. This experimental
science, born in the aftermath of the Second World War and fuelled by
developments in electronics and
computing, seeks answers to fundamental cosmological
problems. Barton was one of the greatest experimentalists, working in deep
underground locations all over world and inspiring a generation of physicists to
follow him.

Born in 1923, he was a wartime student at University College London during its
evacuation to Bangor. Clearly, wartime requirements influenced the curriculum
and he gained a ``Certificate of Proficiency in Radio-Physics'', which was to mark
out his future path. He graduated in 1943, receiving the Granville Prize, annually
awarded to the best Physics graduate of London University. He did his National
Service at Marconi in Chelmsford, Essex, working on military electronics. Here
he acquired a thorough grounding in the basics of electronic design and
construction techniques.
After the war he became a PhD student at Birkbeck College, London University,
working with E.P. George on cosmic rays. His first experiments were performed
at an altitude of 3,457 metres, high on the Jungfraujoch in Switzerland. An
amusing aside in the report thanks the director of the Jungfraujoch railway for
the loan of 15 tonnes of coal that had been used for absorber. This use of
available materials was to become something of a trademark of Barton's
experiments.

In 1954 he began four years at the nascent University College of the West
Indies, Jamaica. As well as his teaching, Barton continued his research,
measuring the cosmic muon flux as a function of depth in the sea. For this he
built a transistorised cosmic-ray ``telescope'' using Geiger tubes contained in a
pressure vessel, which was lowered to depths of 3,000m. Calibrations were
made underground at Norton Hill Colliery in Somerset which were Barton's first
experiments in a mine. To record the data he used his own design of tape
recorder, capable of recording eight tracks of digital data across standard
quarter-inch audio tape. This was the first ever use of digital recording on a
scientific experiment. 

In 1958, Barton returned to London and became lecturer at Northern
Polytechnic, almost immediately publishing his first paper describing a
cosmic-ray detector using photomultipliers, which had just become
commercially available. Photomultipliers are vacuum tubes which detect faint
flashes of light, and their large area of sensitivity and high gain meant that large,
robust particle detectors could be built. They are still found at the heart of
innumerable physics experiments and huge numbers are used in medical
imaging. Barton rapidly became one of the world experts in photomultipliers and
their applications. He also had an almost intuitive feel for the ``non-imaging''
optics needed to carry light to the photomultipliers, and used to say to his
students, ``Light doesn't go down a funnel like water does'', when they came up
with ideas that didn't work.

He began a series of experiments to determine the nature of cosmic rays that
could penetrate deep underground. These experiments were performed in the
``Holborn Laboratory'', a series of rooms deep in Holborn Underground station. A
spare platform at Holborn had been converted to offices during the war.
Immediately after the war it was used as a staff hostel, and later many of the
rooms were used by physicists for experiments needing a deep location.
The laboratory rooms were reached through a service door on one of the
Piccadilly Line platforms. They were linked by an extremely narrow corridor, only
wide enough for a single person, running along the edge of what had once been
the platform. It was a dry and dusty environment and there were occasional
problems caused by rodents chewing cables, but it was none the less an
extraordinarily convenient site to work. For many measurements Holborn was
not deep enough and Barton and his colleagues also ran experiments in
Tilmanstone Colliery in Kent and later in the Woodhead Tunnel, a disused
railway tunnel under the Pennines.

In the early 1980s Barton started on a series of studies on meteorites. His low
background laboratory was ideal for identifying trace radioisotopes produced in
the meteorites in space before they hit the earth. This work led on to a search for
``superheavy elements''. Theoretical analysis suggested that while nuclides
heavier than Uranium were unstable, there would be an ``island of stability''
around element 114 which would have half-lives long enough to exist in nature.
Others had already undertaken searches in a range of samples, particularly
meteorites, and some had claimed positive results.
Together with a group from Leeds, Barton repeated the experiments and,
despite having more sensitive equipment, saw no superheavy elements. Years
later, element 114 was made artificially at Darmstadt and was found to have a
half-life of 30 seconds, a full 15 orders of magnitude smaller than the original
predictions. For Barton this was a vindication of his belief that theoretical
predictions must be tested by experiment and that theoreticians are often just
plain wrong.

When the Physics Department of what was now the Polytechnic of North London
closed in 1984, Barton officially retired, devoting himself to research. He held
honorary posts at Birkbeck and at Queen Mary, London University. In 1993 the
Holborn Laboratory was closed, following increasing safety concerns in the
wake of the King's Cross fire. Barton transferred his underground laboratory to
the Eisenhower Centre, a wartime control centre near Goodge Street, and, when
the lease on this expired, to the basement at Queen Mary, not really deep
enough but workable. Increasing frailty did not deter him -- an ingenious
assembly of car jacks enabled him single-handed to move several tonnes of
lead shielding, no mean feat in a cramped laboratory packed with chemicals,
electronics, computers, domestic appliances such as freezers and all the latest
state-of-the-art instruments that he could get his hands on.

He became a member of the team that built the Sudbury Neutrino Observatory
(SNO) deep in a Canadian nickel mine. The observatory relies heavily on the
use of photomultiplier tubes in a harsh environment that must be as free as
possible from radioactive contamination. Barton's contributions were pivotal and
without them SNO's evidence that neutrinos emitted by the Sun change their
``flavour'' on their way to the Earth would be much less convincing.
In 1988, Neil Spooner, then at Imperial College, London, and Professor Peter
Smith of the Rutherford Laboratory were forming a new collaboration to study
dark matter and to hunt for the elusive ``WIMPS'' (weakly interacting massive
particles). They knew that they needed a deep site to shield the detectors from
cosmic rays and that the Boulby potash mine in North Yorkshire was the
deepest mine in Britain. Barton was enthusiastic about this new project and
went with Spooner on the site visit to help persuade the mine management to
accept an underground laboratory. They were successful and the Boulby Dark
Matter Collaboration came into being, operating a range of dark matter detectors
in the rocksalt seams, one kilometre underground. While he never visited the
site again, Barton continued to make vital contributions to the collaboration. The
new surface building there has recently been named the John Barton Building.

Much of Barton's pioneering work was made with relatively cheap equipment he
built himself, using the very latest technology available to him, but always on a
shoestring budget. Once, when asked why he never applied for grants from the
Science Research Council, he replied that, if one applied for a grant, then one
had to write reports on the grant, irrespective of the scientific results, and that
then one ended up believing what one had written in the reports.
Most of his career was spent at Northern Polytechnic, in a period when
polytechnics rarely did fundamental research. Barton managed to, despite the
environment. In 1968 he put together a pack of 50 research papers and
submitted them for a DSc at London University, because he wanted to show that
it was possible to do good science in a polytechnic with a supportive head of
department.

John Barton was a shy, private and unassuming man but you knew within the
first minute of meeting him that you were in the presence of an exceptionally
talented and intelligent person. He was an enthusiastic walker and always took
an annual walking holiday, most recently a strenuous traverse of Corsica. He
loved the cinema and chose to live in Hampstead because of the proximity to the
Everyman Cinema. Barton worked in his laboratory almost daily until his final
illness. On becoming housebound, he bought the first television he had ever
owned, typically finding even Bang \& Olufsen's superior specification left much
to be desired.\\

\noindent\emph{John Clifford Barton, physicist: London 29 September 1923; 
Senior Lecturer in Physics, University College of the West Indies 1954-58; 
Lecturer in Physics, Northern Polytechnic (later Polytechnic of North London) 1958-84, 
Head of Physics Department 1971-84; died London 14 October 2002. }\\

\noindent John McMillan \\
\\
\noindent This obituary first appeared in the daily newspaper ``The 
Independant'',\\
London, 29th November 2002.\\
\\ 
j.e.mcmillan@sheffield.ac.uk\\
\\
Department of Physics and Astronomy, \\
The University of Sheffield, Sheffield, S3 7RH, Great Britain.\\

\pagebreak

\nocite{*}
\bibliographystyle{unsrt}
\bibliography{barton}
\end{document}